\documentclass[12pt]{article}

\usepackage{graphicx}
\usepackage{epstopdf}
\usepackage[body={17.5cm, 21cm},right=2cm]{geometry}
\usepackage{amssymb}
\usepackage{amsmath}
\usepackage{psfrag}
\usepackage{epsfig}
\usepackage{cancel}

\usepackage[all]{xy}
\usepackage{feynmf}

\newcommand{\bra}{\langle}
\newcommand{\ket}{\rangle}
\newcommand{\cH}{{\cal H}}

\newcommand{\x}{{\bf x}}

\newcommand{\y}{{\bf y}}
\newcommand{\p}{{\bf p}}
\newcommand{\q}{{\bf q}}
\newcommand{\kk}{{\bf k}}
\newcommand{\ad}{a^\dagger}
\newcommand{\kp}{{\bf k'}}

\newcommand{\pb}{\overline{P}}
\newcommand{\rb}{\overline{R}}
\newcommand{\rr}{{\bf r}}

\newcommand{\z}{{\bf z}}

\newcommand{\tad}{\begin{picture}(30,30)(-15,5)\thicklines
\put(0,12){\circle{25}}
\put(0,0){\circle{5}}
\end{picture}}

\newcommand{\tadp}{\begin{picture}(30,30)(-15,5)\thicklines
\put(0,12){\circle{25}}
\put(0,0){\circle*{5}}
\end{picture}}

\newcommand{\myloop}{\begin{picture}(30,30)(-15,5)\thicklines
\put(0,12){\circle{25}}
\put(0,0){\circle{5}}
\put(0,24){\circle{5}}
\end{picture}}

\newcommand{\looppr}{\begin{picture}(30,30)(-15,5)\thicklines
\put(0,12){\circle{25}}
\put(0,0){\circle*{5}}
\put(0,24){\circle{5}}
\end{picture}}

\newcommand{\looppp}{\begin{picture}(30,30)(-15,5)\thicklines
\put(0,12){\circle{25}}
\put(0,0){\circle*{5}}
\put(0,24){\circle*{5}}
\end{picture}}

\newcommand{\loopprr}{\begin{picture}(30,30)(-15,5)\thicklines
\put(0,12){\circle{25}}
\put(0,0){\circle*{5}}
\put(-10,19){\circle{5}}
\put(10,19){\circle{5}}
\end{picture}}

\newcommand{\loopppp}{\begin{picture}(30,30)(-15,5)\thicklines
\put(0,12){\circle{25}}
\put(0,0){\circle*{5}}
\put(-10,19){\circle*{5}}
\put(10,19){\circle*{5}}
\end{picture}}

\newcommand{\loopppr}{\begin{picture}(30,30)(-15,5)\thicklines
\put(0,12){\circle{25}}
\put(0,0){\circle*{5}}
\put(-10,19){\circle*{5}}
\put(10,19){\circle{5}}
\end{picture}}

\newcommand{\loopprp}{\begin{picture}(30,30)(-15,5)\thicklines
\put(0,12){\circle{25}}
\put(0,0){\circle*{5}}
\put(-10,19){\circle{5}}
\put(10,19){\circle*{5}}
\end{picture}}

\newcommand{\looppn}{\begin{picture}(30,30)(-15,5)\thicklines
\put(0,12){\circle{25}}
\put(0,0){\circle*{5}}
\put(-10,6){\circle*{5}}
\put(10,6){\circle*{5}}
\put(-10,19){\circle*{5}}
\put(10,19){\circle*{5}}
\put(0,24){\circle*{2}}
\put(5,23){\circle*{2}}
\put(-5,23){\circle*{2}}
\end{picture}}

\newcommand{\looppnw}{\begin{picture}(30,30)(-15,5)\thicklines
\put(0,12){\circle{25}}
\put(0,0){\circle{5}}
\put(-10,6){\circle*{5}}
\put(10,6){\circle*{5}}
\put(-10,19){\circle*{5}}
\put(10,19){\circle*{5}}
\put(0,24){\circle*{2}}
\put(5,23){\circle*{2}}
\put(-5,23){\circle*{2}}
\end{picture}}

\newcommand{\twopoint}{
\begin{picture}(40,30)(-5,0)\thicklines
\put(0,0){\circle*{5}}
\put(0,0){\line(1,0){30}}
\put(30,0){\circle*{5}}
\end{picture}}

\newcommand{\threepoint}{\begin{picture}(70,30)(-5,0)\thicklines
\put(0,0){\circle*{5}}
\put(0,0){\line(1,0){60}}
\put(30,0){\circle{5}}
\put(60,0){\circle*{5}}
\end{picture}}

\newcommand{\fourpoint}{\begin{picture}(100,30)(-5,0)\thicklines
\put(0,0){\circle*{5}}
\put(0,0){\line(1,0){90}}
\put(30,0){\circle{5}}
\put(60,0){\circle{5}}
\put(90,0){\circle*{5}}
\end{picture}}

\newcommand{\lineppp}{\begin{picture}(100,30)(-5,0)\thicklines
\put(0,0){\circle*{5}}
\put(0,0){\line(1,0){60}}
\put(30,0){\circle*{5}}
\put(60,0){\circle*{5}}
\end{picture}}

\newcommand{\linepprp}{\begin{picture}(100,30)(-5,0)\thicklines
\put(0,0){\circle*{5}}
\put(0,0){\line(1,0){90}}
\put(30,0){\circle*{5}}
\put(60,0){\circle{5}}
\put(90,0){\circle*{5}}
\end{picture}}

\newcommand{\lineprprp}{\begin{picture}(100,30)(-5,0)\thicklines
\put(0,0){\line(1,0){120}}
\put(0,0){\circle*{5}}
\put(30,0){\circle{5}}
\put(60,0){\circle*{5}}
\put(90,0){\circle{5}}
\put(120,0){\circle*{5}}
\end{picture}}

\begin{document}

\setcounter{page}{0}
\thispagestyle{empty}

\begin{titlepage}

\begin{center}

~

\vspace{1.cm}

{\LARGE \sc{
Renormalized Thermal Entropy \\[2mm]
in Field Theory
}}\\[1cm]
{\large Sergio Cacciatori$^{\rm a}$, Fabio Costa$^{\rm b,c}$ and Federico Piazza$^{\rm c}$,
\\[0.6cm]

\vspace{.2cm}
{\small \textit{$^{\rm a}$
Dipartimento di Scienze Fisiche e Matematiche, \\ Universit\`a dell'Insubria, via Valleggio 11, I-22100 Como}}

\vspace{.2cm}
{\small \textit{$^{\rm b}$
Institut f\"ur Quantenoptik und Quanteninformation (IQOQI), \"Osterreichische Akademie der Wissenschaften,
Boltzmanngasse 3, A-1090 Wien, Austria}}}

{\small \textit{$^{\rm c}$
Perimeter Institute for Theoretical Physics\\
Waterloo, Ontario, N2L 2Y5, Canada}}

\end{center}

\begin{abstract}

Standard entropy calculations in quantum field theory, when applied to a
subsystem of definite volume, exhibit area-dependent UV divergences that
make a thermodynamic interpretation troublesome. In this paper we
define a renormalized entropy which is related with the Newton-Wigner
position operator.
Accordingly, whenever we trace over a region of space, we trace away
degrees of freedom that are localized according to Newton-Wigner localization but not in
the usual sense.
We consider a free scalar field in d+1 spacetime dimensions prepared in
a thermal state and we show that our entropy is free
of divergences and has a perfectly sound thermodynamic behavior. In the
high temperature/big volume limit our results agree with
the standard QFT calculations once
the divergent contributions are subtracted from the latter. In the limit of low
temperature/small volume the entropy goes to zero but with a different
dependence on the temperature.

\end{abstract}
\end{titlepage}

\section{Introduction}

Thermodynamics is a very powerful tool for describing complex physical systems. Beside its evident experimental success in the laboratory,
thermodynamics is used in the everyday practice of cosmology, for instance when the conservation of entropy is applied to a comoving volume of the
expanding Universe (see e.g. \cite{kolb}); on a more speculative level, the quest for a theory of quantum gravity often makes use of thermodynamic
and entropic arguments, since they are supposed to be somewhat independent of the details of the underlying dynamics.
Especially in view of such adventurous applications, it is certainly worth understanding better and better the connection between thermodynamics and
microphysics within those physical regimes which are best known and under control. The tendency to thermal equilibrium, that in classical statistical
physics relies on ergodic or mixing hypotheses, in quantum mechanics seems to be naturally driven by the correlations that a subsystem inevitably
develops with its environment.
This general view, which has occasionally appeared in the literature at the level of common wisdom, is now being put on firmer ground e.g. in the
interesting book \cite{gemmer} and in related ongoing works.
During time evolution, the reduced quantum state of a subsystem tends in fact to approach that of a thermal Gibbs state
$\rho_{\rm thermal} \propto e^{- \beta H}$, $H$ being the Hamiltonian operator of the subsystem. This happens, under some generic circumstances and
on time scales which are thoroughly discussed in \cite{gemmer}, even if the initial state of the subsystem is very different from $\rho_{\rm thermal}$,
e.g. in the case when the entire system is initially prepared in a product state $|{\rm subsystem}\ket \otimes |{\rm environment}\ket$.

Our most successful microscopic description of physical interactions, quantum field theory (QFT), faces some difficulties when asked to reproduce
coarse-grained meaningful thermodynamic quantities. In particular, as first noted in \cite{wilc2}, the UV-divergencies encountered in the calculation
of entropy are of  a relatively uncommon type. If a finite system is in a thermal state, its  entropy can be calculated with standard methods  giving
a thermodynamically sound result (see e.g. \cite{kapusta}). For the reasons described above, however, it is also interesting to consider, instead of
the entire system, a subsystem occupying a finite portion of the entire volume. In this case, the entropy exhibits a UV-divergent ``vacuum" contribution
proportional to the boundary of the subsystem.

\subsection{Thermal Entropy and UV Divergences}

To be more definite, consider a system $S$ whose dynamics is described by a QFT Hamiltonian $H$ and put it in a thermal state $\rho_{\rm total}\propto e^{-\beta H}$. Then consider a region of space $P$ ($P$ stands for ``place") of definite volume inside $S$ and call $R$ the rest of the system.
The state in $P$ is obtained by tracing out the irrelevant degrees of freedom belonging to $R$,
$\rho = Tr_R \rho_{\rm total}$. Then calculate the Von Neumann entropy
$S=-Tr\rho\ln\rho$, which is the appropriate generalization of thermodynamical entropy for generic quantum states (see e.g. \cite{wehrl,landau}).
Schematically, in four spacetime dimensions, one finds
\begin{equation} \label{s}
S\, =\, S_{\rm vac}(A, \Lambda) \, +\,  S_{\rm therm}(V, T).
\end{equation}
Here $A$ and $V$ are the boundary area and the volume of $P$ respectively, $\Lambda$ is a UV cutoff and $T=1/\beta$ is the temperature. The term $S_{\rm vac}$ is the UV-divergent entanglement entropy of the vacuum (see, among others \cite{sorkin,sre,callan,casini2}), obtained with the same procedure in the limit of zero temperature, {\it i.e.} when $\rho_{\rm total} = |0\ket \bra0|$.
The general form of $S_{\rm vac}$ is \cite{casini}
\begin{equation} \label{structure}
S_{\rm vac} \, =\, c\,  \Lambda^2 A + {\cal O}( \Lambda^2  A)^{1/2}\, ,
\end{equation}
$c$ being a regularization dependent number of order one, for
more general boundaries see, e.g. \cite{fursaev}.
The expansion \eqref{structure} follows quite generally e.g. from the heat kernel methods used in \cite{callan}. The finite component $S_{\rm therm}(V, T)$, on the other hand, is  the meaningful thermodynamic quantity; for a massless field it typically scales as
$\sim V T^3$ in the big volume/high temperature limit. A few comments are now in order.

While the leading area-dependent vacuum divergence \eqref{structure} can be checked in a variety of ways, bringing out the subleading finite term $S_{\rm therm}$ is not trivial and, to our knowledge, has been done explicitly only for conformal field theories. In 1+1 dimensions Calabrese and Cardy  \cite{cala}, by exploiting the analytic properties of the theory, found a structure of the type \eqref{s} with a term $S_{\rm vac}$ logarithmically divergent with the cutoff. By using insights from AdS/CFT correspondence, Ryu and Takayanagi have been able to extend the result to higher dimensions\footnote{In theories with AdS gravity duals, the thermal entropy \eqref{s} is  conjectured to be proportional to the area of a suitably defined hypersurface living in a AdS black hole bulk. The two contributions in eq. \eqref{s} have then two clear and distinct origins: the portion of hypersurface closest to the boundary of AdS accounts for the divergent``vacuum'' term; the finite part is !
 due to the bending of the hypersurface
deeper in the bulk, due to the presence of the central black hole.} \cite{ads}. However, it would be surprising if thermal entropy did not have the structure \eqref{s} in general. The entanglement of the vacuum is in fact a UV effect and should be there also for generic finite energy states. The two separate terms in \eqref{s} are thus expected in any plausible QFT theory where the highest energy modes decouple from the low energy physics. Of course,
if thermal entropy did not have the form \eqref{s} and, say, some divergent term were also temperature dependent, then the intent of deriving meaningful coarse-grained quantities would be even more troublesome.

One may object that the volume dependent term in \eqref{s} wins over the area-term in the thermodynamic limit. In thermodynamics, however, volumes have to be big
in comparison, say, to the typical distances between particles. On the other hand, by taking for instance a typical cosmological setup ($T^2 \sim M_{Pl} \times {\rm Hubble}$) and $\Lambda \sim M_{Pl}$, it is easy to see that the thermodynamic term overcomes in \eqref{s} only for volumes much bigger than the Hubble scale! We expect thermodynamics to be applicable in much less extreme conditions.
One may also object that, rather than the absolute value, only entropy differences are meaningful. Still, the area dependent term in \eqref{s} spoils the attempts of a thermodynamic description for subsystems whose size is (adiabatically) changing in time. In this regard, once again, a comoving volume in an expanding Universe is perhaps the cleanest example.

Due to the non trivial dependence of the divergence on area, the quantity \eqref{s} cannot be renormalized by standard methods, \emph{i.e.} by adding local counterterms to the Lagrangian. Moreover, since the result \eqref{structure} has been carried out for free and conformal theories, we are bound to have divergences regardless of the asymptotic behavior of the couplings or the UV completion of the theory, as long as such a completion is still a field theory. Of course, as proposed in \cite{wilc2}, one can always \emph{subtract} the divergent terms. The latter, however, are not more ``spurious" than the widely accepted entanglement of the vacuum\footnote{It is sometimes emphasized that the Von Neumann entropy of a subsystem has a different meaning depending on whether or not the entire system is in a pure state. Thus, one may be tempted to take the entanglement of the vacuum seriously \cite{reznik} and just reject the divergent term
$S_{\rm vac}$ in \eqref{s} as spurious. Consider, however, the entire Universe
in a pure state; we can expect thermalization to occur over some region $S$ after local equilibrium is reached {\it i.e.} $\rho_S \simeq e^{-\beta H}$. Deep inside $S$ one can eventually consider the subsystem $P$. The Von Neumann entropy of $P$ is clearly both a thermodynamic entropy -- because $S$ is in a nearly thermal state -- and an entanglement entropy -- because the whole Universe is in a pure state. Note also that, starting from the entangled vacuum of QFT, one can always construct states which are less and less entangled; such states, although probably not generic nor thermodynamically very interesting, would end up having a negative entropy after the subtraction of $S_{\rm vac}$.}. In plain contradiction with the general view/common wisdom illustrated at the beginning of this introduction, equation \eqref{s} is just saying that the state of a generic subsystem in QFT is actually very far from being thermal!

Since the Hamiltonian operator $H$ is an integral of a local density, one may expect that a state of the form $e^{-\beta H}$ would factorize over contiguous regions of space giving $e^{-\beta H_P}\otimes e^{-\beta H_R}$, where $H_P$ and $H_R$ are the integrals of Hamiltonian density extended only to subsystems $P$ and $R$ respectively. If this were the case, tracing over $R$ would trivially give a thermal state in $P$. However,
the energy $H$ hides a relevant amount of non-extensiveness that doesn't allow this factorization. The ``inside" and ``outside" contribution, $H_P$ and $H_R$, do not add up to the total Hamiltonian because of the UV-divergent contact term $H_I$ coming from the gradients across the boundary between $P$ and $R$. In QFT, because of the singular nature of the interaction $H_I$ between $P$ and its environment $R$, the general arguments of \cite{gemmer} are not applicable.

\subsection{Renormalizing the Entropy}

The above difficulties can be ascribed to an inconvenient choice of degrees of freedom. To see what that means, note that the system/region of space in question has two complementary descriptions \cite{fedo,fedo2,pc}.
In compliance to common intuition, $P$ is described
in classical general relativity by a subset $\pb$ -- more specifically,
a submanifold -- of the points/events at a given time-like coordinate $t$.
 On the other hand, as a quantum subsystem, $P$
is described by a Hilbert space $\cH_P$, which is a factor in the tensor product
decomposition
($\cH = \cH_P\otimes \cH_R$) of the total Hilbert space of the field theory under examination\footnote{Note that partitioning a quantum system \cite{paolo,zan2} is more subtle than making the partition of a set.
While a finite set admits a finite number of possible partitions, a finite number of quantum degrees of freedom can be divided in an infinite number of inequivalent ways. The simplest example is a four dimensional quantum system, $\cH = \mathbb{C}^4$, whose partitions into ``two spins'', $\simeq \mathbb{C}^2\otimes \mathbb{C}^2 $, are in one to one correspondence with the elements of the infinite group $SU(4)/SU(2)^2$ \cite{pc}.}.
In this paper we show that if we associate the region
of space $\pb$ with a more suitable -- although unconventional -- set of
quantum degrees of freedom $\cH_P$, the entropy is already "renormalized" and has a perfectly sound thermodynamic behaviour. For a massless free scalar in the large volume/high temperature limit we find $S = c(d) V T^d$, where $d+1$ is the dimension of
spacetime and $c(d)$ a numerical factor that we calculate explicitly (see eqs. \eqref{hightentropy}). In this limit our results are consistent with those obtained by \cite{cala,ads} once their divergent contribution is subtracted. However, there is no trace of any area-dependent term in our calculations and no infinities are encountered except, of the IR type, in the $1+1$ dimensional massless case in the limit $T\rightarrow0$. In the same limit, our results differ from the ``finite piece" $S_{\rm therm}$ that is found in \eqref{s} by using the conventional approach, although entropies tend to zero in both cases.
For $T\rightarrow0$ we find in fact $S\sim - V T^d \ln(V T^d)$, while \cite{cala,ads} find $S_{\rm therm} \sim
(V T^d)^{(d+1)/d}$. We comment on this in the conclusions.

Before describing our calculation in section \ref{sketches} it is perhaps worth spending few more words and see in which sense our anomalous renormalization procedure underlies a different ``localization scheme" (see \cite{pc} for more details). The main point here is how to pick a bunch of (local) quantum degrees of freedom $\cH_P$ out of a larger system.
A tensor product structure (TPS) -- or quantum partition -- can be assigned by specifying the observables of the individual subsystems \cite{zan2,liu}.
In a composite system $\cH_P \otimes \cH_R$ two sets of observables ${\cal A}^j(P)$ and ${\cal A}^k(R)$, separately defined in $P$ and $R$ respectively, commute by construction:
\begin{equation}\label{commute}
[ {\cal A}^j(P), {\cal A}^k(R)]\, =\, 0\qquad \text{for any}\ \  j,k.
\end{equation}
The point here is that such a trivial result can be applied the other way around \cite{zan2}:
if,  within the algebra of observables acting on $\cH$, we manage to isolate two commuting subalgebras ${\cal A}^j(P)$ and ${\cal A}^k(R)$, they induce a unique\footnote{Actually, only if
the two subalgebras generate the entire algebra of operators on $\cH$ \cite{zan2}}
bipartition $\cH = \cH_P \otimes \cH_R$ on the whole system.
In the conventional calculation of entropy \eqref{s} it is implicitly
assumed that the quantum degrees of freedom $\cH_P$ of
a region of space $\pb$ at time $t$ are those defined by the set of local
relativistic fields $\phi(t, \x \in \pb)$ and their conjugate momenta
$\pi(t, \x \in \pb)$. In fact, thanks to the canonical commutation relations, the two subalgebras generated by $\phi$ and $\pi$ with labels $\x$ inside and outside $P$ satisfy \eqref{commute}, and therefore induce a TPS. Such a TPS is the conventional localization scheme in QFT.

In order to renormalize the entropy, in this paper we use an alternative set of commuting operators  -- and their corresponding TPS -- as a new rationale to isolate the quantum degrees of freedom of
$P$. We consider a free scalar field in $d+1$ dimensional Minkowski spacetime. The normal ordered Hamiltonian reads
\begin{equation} \label{hamiltonian}
H\ =\ \int d^d k \, w_k \, \ad_\kk\, a_\kk,
\end{equation}
where $w_k = \sqrt{{\bf k}^2 + m^2}$
and operators $a_\kk$ satisfy
the commutation relations $[a_\kk, a_\kp]=0,\,  [a_\kk, \ad_\kp]=\delta^3(\kk - \kp)$.
Instead of the relativistic fields and their conjugate momenta we introduce the
``Newton-Wigner" fields $a(\x)$ just as the Fourier transforms of $a_\kk$:
\begin{equation} \label{newton}
a(\x) = \frac{1}{(2 \pi)^{d/2}} \int d^d k \, a_\kk\, e^{ i \kk \cdot \x} , \qquad
\ad(\x) = \frac{1}{(2 \pi)^{d/2}} \int d^d k \, \ad_\kk\, e^{- i \kk \cdot \x} ;
\end{equation}
The above defined operator $\ad(\x)$ is directly related to the Newton-Wigner (NW) position operator \cite{nw} in that, acting on the vacuum, it produces an eigenvector of eigenvalue $\x$. Note that the relativistic invariant measure $1/\sqrt{2 w_k}$ is absent from the integrand and therefore
those operators are not relativistically invariant. This amounts to the fact that a particle perfectly NW-localized according  to some observer is instead ``spread" when described by a boosted one \cite{stefa}. On the other hand, the dynamics is still relativistically invariant because we are not changing the Hamiltonian of the free scalar \eqref{hamiltonian} nor the other generators of the Poincar\'e group. What is usually considered a drawback of the NW approach doesn't worry us too much, particularly here since thermal states with $T>0$ break Lorentz invariance. The obvious reference frame in this setup is the one with four-velocity parallel to the expectation value of the momentum operator, $\bra P^\mu \ket$.  We refer to the extensive literature for more technical details (e.g. \cite{nw,stefa,margaret}) and philosophical implications \cite{flem} of NW operators. A very introductory comparison between the two localization schemes is found in \cite{pc}.

\section{Sketches of the Calculation and Main Results} \label{sketches}

Since the entire problem is stationary, we can forget about the time coordinate. At some given time we cut a $d+1$ dimensional Minkowski space into two connected regions: $\pb$, of finite volume, and $\rb$, {\it i.e.} $\pb \cup \rb = \mathbb{R}^3$. Our results are independent of the shape of $\pb$.
We distinguish spatial coordinates belonging to different regions using labels $\p$, $\p '$, $\p_j$ $\ldots$ for points inside $\pb$, $\rr$, $\rr '$, $\rr_j$ $\ldots$ for those in $\rb$ and $\x$, $\y$ $\ldots$ for generic points in ${\mathbb R}^d$.
One of the basic properties of NW localization is that the vacuum of the theory is a product state, {\it i.e.} $|0\ket = |0_P\ket \otimes|0_R\ket$ \cite{pc}. Moreover, starting from the vacuum, we can repeatedly apply $\ad(\x \in \pb)$ and $\ad(\x \in \rb)$ and generate two independent Fock spaces:
\begin{equation}
\cH_P =  {\mathbb C} \oplus P_1 \oplus P_2 \oplus \ldots \oplus P_n\oplus\ldots \;\qquad
\cH_R =  {\mathbb C} \oplus R_1 \oplus R_2 \oplus \ldots \oplus R_n\oplus\ldots \ .
\end{equation}
This Fock space decomposition of the regions is the distinctive feature
of NW localization and allows an intuitive representation of particles
localized in space. This is impossible in the standard localization, as
the Reeh-Schlieder theorem forbids the existence of particle states of
finite energy. The most striking
consequence is that the vacuum is here a product state, while it is
entangled in the standard scheme.
In each Fock sub-space of given particle number we choose the obvious basis
\begin{equation} \label{basis}
P_n\, \rightarrow \, |\p_1 \dots \p_n\ket = \frac{1}{\sqrt{n!}} \ad(\p_1) \dots \ad(\p_n)|0\ket \qquad
R_n\, \rightarrow \, |\rr_1 \dots \rr_n\ket = \frac{1}{\sqrt{n!}} \ad(\rr_1) \dots \ad(\rr_n)|0\ket .
\end{equation}
When we calculate traces we therefore sum on a basis in a Fock space of given particle number and then sum over all Fock subspaces. For example, if we restrict to block diagonal states (as are all those appearing here), we have
\begin{equation}
{\rm Tr}_R\,  \cdot = \bra 0_R| \cdot |0_R\ket + \int_R d^d r  \bra\rr| \cdot |\rr\ket +
\int_R d^d r  d^d r' \bra\rr \rr'| \cdot |\rr \rr'\ket +\dots
\end{equation}
(here and in the following we write $P$, $R$,  for $\pb$, $\rb$, understanding the identification of regions with subsystems via the NW scheme).

We put the entire system in a (non normalized) thermal state, $\rho_{\rm total} = e^{- \beta H}$, where $H$ is defined in \eqref{hamiltonian}. We call $\rho$ the reduced state in $P$: $\rho \equiv {\rm Tr}_R \rho_{\rm total}$ and we calculate Von Neumann entropy by means of the formula \cite{callan}
\begin{equation} \label{entropy}
S \, \equiv\,  -{\rm Tr}_P (\rho \ln \rho)\, = \, \left. \left(-\frac{d}{d n} + 1\right)\ln {\rm Tr}_P \, \rho^n \right|_{n=1}
\end{equation}
which allows us to lose track of normalization factors.

It's easy to check that $\rho$ (as well as $\rho_{\rm total} $) is block diagonal on the particle number subspaces of its Fock space. Its matrix elements on each subspace are expressible in terms of the crucial two point function defined in the one-particle sector
\begin{equation}
K(\p, \p')\, \equiv \frac{\bra \p|\, \rho\, |\p'\ket}{\Omega},
\end{equation}
where $\Omega \equiv \bra 0_P|\, \rho \, |0_P\ket$ is the vacuum-vacuum matrix element.
The function $K$ is itself an infinite series obtained by tracing over $R$ (see eq. \eqref{KK} below)
and can be nicely written (see eq. \eqref{twopoint}) in the diagrammatic formalism to be introduced in the next section. Since we are dealing with a free theory, the generic matrix element $\bra\p_1 \p_2 \dots \p_n| \rho|\p_1' \p_2' \dots \p_n'\ket $ is expressed as a combination of products of $K$s in equation \eqref{term}.

Since $\rho$ is an operator acting on $P$, when calculating matrix elements of $\rho^n$, integrations over the variables $\p$ have to be carried out in each Fock subspace. A final integration over the same variables has to be done in order to obtain ${\rm Tr}_P \rho^n$. The terms in the corresponding series rearrange (see eq. \eqref{rearrange}) and this number can be written in closed form, again, in terms of the two point function $K$ as
\begin{equation} \label{drops}
{\rm Tr}_P \, \rho^n \ = \Omega^n \ \exp\left(\sum_{j=1}^{\infty}\frac{1}{j} {\rm Tr}\, K^{j \cdot n}\right)\, ,
\end{equation}
where, by definition,
\begin{equation}
{\rm Tr}\, K^m \ \equiv \ \int_P d^d p_1 d^d p_2 \dots d^d p_m\, K(\p_1,\p_2)K(\p_2,\p_3) \dots K(\p_m,\p_1).
 \end{equation}
Note that the normalization factor $\Omega^n$ in \eqref{drops} drops when used in formula
\eqref{entropy}. Note also that each factor $K$ in the above integral is itself a series of integrals over the $\rr$ variables.
A consistent part of this work is finally devoted to evaluate the above quantity in the high temperature,  $V/\beta^d \rightarrow \infty$, and small temperature, $V/\beta^d \rightarrow 0$, limits. Two distinct behaviors of ${\rm Tr}\, K^n$ as a function of $n$ follow.

In the high temperature limit the leading term of the series giving ${\rm Tr}\, K^n$ is the one containing only integrations over the $\p$ variables: the other terms converge and give a subleading contribution. The actual proof of that is rather involved and part of it appears in the Appendix.
The only integrations left are those inside $P$. Those are all of the same order in $V T^d$, although they scale as $1/n^d$, where $n$ is the number of integration variables. The corresponding behavior of  ${\rm Tr}\, K^n$, when used in \eqref{drops} and \eqref{entropy},  gives an extensive entropy. In this limit, apart from numerical factors to be found in eqs. \eqref{tracciakn} and \eqref{hightentropy}, one finds in fact ($1/VT^d \rightarrow 0$)
\begin{equation}
{\rm Tr}\, K^n \, \simeq \, \frac{V T^d}{n^d}\,  +\,  {\cal O}(1/V T^d)^0, \quad \qquad \quad
S\,  \simeq \, VT^d\,  +\,  {\cal O}(1/V T^d)^0.
\end{equation}

In the low temperature limit the external integrals are no longer negligible and they have to be summed up. The corresponding series can be explicitly calculated at leading order in $VT^d$. On the other hand, terms with a higher number of internal integrations are subleading and this gives
($VT^d \rightarrow 0$)
\begin{equation} \label{masslesss}
 {\rm Tr}\, K^n \, \simeq \, (V T^d)^n\,  +\,  {\cal O}(V T^d)^{n+1}, \quad \qquad \quad
S\,  \simeq \, - VT^d \ln VT^d \,  +\,  {\cal O}(V T^d).
\end{equation}
The numerical factors are found in eqs. \eqref{tracciaklow} and \eqref{lowent} below. In section \ref{lowmassive}, we will also consider massive fields; in the low temperature limit they behave like \eqref{masslesss} except that the quantity $VT^d$ is each time suppressed by the factor $e^{- m/T}$ (see eq. \eqref{lowmassived1} below).

\section{Formalism and Diagrammatic}

After the general setting described above (basically, eqs. \eqref{hamiltonian}--\eqref{basis}) we give here more details of the calculation.
First of all, we need to calculate the matrix elements of $\rho_{\rm total} = e^{-\beta H}$ in position space; using the basis vectors \eqref{basis} and going to Fourier space, we have
\begin{equation} \label{element}
\bra \x_1 \dots \x_n|\, e^{-\beta H}\,
|\x'_1 \dots \x'_n \ket \, =\, \frac{1}{m! (n-m)!}
\sum_{\sigma\in S_n} I_\beta(\x_1 - \x'_{\sigma(1)})\cdots I_\beta(\x_n - \x'_{\sigma (n)}).
\end{equation}
Here, of the total $n$ points, $m$ is the number of points inside $P$, $S_n$ denotes the group of permutations over $n$ elements and
\begin{equation} \label{I}
I_\beta (\x - \x') \equiv \bra \x|\, e^{-\beta H}\,
|\x'\ket \, = \frac{1}{(2 \pi)^d}\int d^d k \, e^{\, i \, \kk \cdot (\x - \x')\, - \, \beta w_k}\ .
\end{equation}

Explicit expressions of the two point function $I$ and its massless limit follow for $d=1$
\begin{equation} \label{1dim}
I_\beta(x) \ = \ \frac{m \beta}{\pi \sqrt{\beta^2 + x^2}}\, K_1(m\sqrt{\beta^2 + x^2}) \ \simeq\
\frac{\beta}{\pi (\beta^2 + x^2)} + {\cal O}(m^2)
\end{equation}
and $d=3$:
\begin{equation} \label{3dim}
I_\beta (\x) \ = \ \frac{m^2 \beta}{2 \pi^2 (\beta^2 + x^2)}\, K_2(m\sqrt{\beta^2 + x^2}) \, \simeq \
\frac{\beta}{\pi^2 (\beta^2 + x^2)^2} + {\cal O}(m^2) .
\end{equation}
Here $K_j$ are the modified Bessel functions of the second kind.
Note that \eqref{I} is not the usual QFT thermal correlator and, as such, it is not periodic in  $\beta$. This reflects the
fact that we are not working in the usual thermic representation, where traces are taken by functional integration over a compactified Euclidean manifold.

A crucial property of the two point function $I_\beta$ that follows straightforwardly from its expression \eqref{I} in Fourier space is
\begin{equation} \label{crucial}
\int d^d z \, I_\beta(\x - \z) I_\gamma(\z - \y)\ = \ I_{\beta + \gamma} (\x - \y).
\end{equation}
By iteration we also have
\begin{equation} \label{iteration}
I_\beta^n(\x - \y) = I_{n\beta}(\x - \y),
\end{equation}
where the $n^{\rm th}$ power of $I_\beta$ has been implicitly defined in an obvious way.

Matrix elements with a different number of particles on the two sides vanish, because in transformation \eqref{newton} the number of creation and annihilation operators is preserved; our matrix $\rho_{\rm total}$ is thus block diagonal in the subspaces of given particle number. The same property is retained by the reduced density matrix $\rho$ with respect to the local Fock space $\cH_P$, so that we only need to calculate the matrix elements $\bra\p_1\dots \p_n| \rho|\p'_1\dots\p'_n\ket$.

First, we define $\Omega$ as the matrix element of $\rho$ on the local vacuum in $P$:
\begin{multline}
\Omega \, \equiv \, \!\bra0_P|e^{-\beta H} |0_P\ket \ =\ \\
= (\bra0_R|\otimes \bra0_P|)\, \rho\, (|0_P\ket \otimes|0_R\ket) \ +
\int_R d^d r  (\bra \rr | \otimes \bra 0_P|)e^{-\beta H}
(|0_P \ket \otimes |\rr\ket) +\\
\int_{R\times R} d^d r_1 d^d r_2 \, (\bra\rr_1 \rr_2|\otimes \bra0_P|)\, e^{-\beta H}\,
(|0 \ket_P \otimes |\rr_1 \rr_2\ket_P) +\ \dots
\end{multline}

In terms of the two point function $I_\beta (\x - \x')$ we have
\begin{multline}\label{terms}
\Omega \ = \ \sum_{n = 0}^{\infty} \frac{1}{n!} \int_{R^n} d^d r_1  \dots d^d r_n\, \sum_{\sigma \in S_n}
\ \prod_{j=1}^{n}\, I_\beta (\rr_j - \rr_{\sigma(j)}) \ =
\\
1\ +\int_{R} d^d r\, I_\beta(\rr - \rr)\ +\ \frac{1}{2} \int_{R\times R} d^d r_1 d^d r_2\, \left[  I_\beta(\rr_1 - \rr_1) I_\beta(\rr_2 - \rr_2) +
 I_\beta(\rr_1 - \rr_2)  I_\beta(\rr_1 - \rr_2)\right] \\ \ +\, \dots
\end{multline}
We can write this kind of expressions in a diagrammatic form; in this way, the vacuum expectation value is given by the sum of all the ``bubble diagrams'':
\begin{equation}
\Omega\ =\ 1\ +
\tad
+\frac{1}{2}\left(\tad \tad + \myloop  \right)+ \dots
\end{equation}
Here and in the following empty circles $\circ$ are points in $R$ and full circles $\bullet$ points in
$P$, lines are the two point function $I$ and two lines getting at the same circle imply integration.
The $n^{\rm th}$ term of the series \eqref{terms} is obtained diagrammatically by taking $n$ empty circles and
connecting them with each other in all possible ways such that each circle is reached by two lines.

The matrix element of $\rho$ living in the one particle sector is a two point function:
\begin{equation} \label{KK}
\bra\p| \rho|\p'\ket\, =\, I_\beta(\p - \p') \, +\, \int_R d^d r\,
\left[I_\beta(\p - \p')I_\beta(\rr - \rr)+ I_\beta(\p - \rr)I_\beta(\rr - \p')\right]\, +\ldots\,
\end{equation}
diagrammatically,
\begin{multline}
\bra\p| \rho|\p'\ket\ =\
\twopoint \ +\ \left(\twopoint \tad \ +\
\threepoint \right)+\\
\frac{1}{2}\left(\twopoint \tad \tad + 2
\threepoint \tad + 2 \fourpoint + \twopoint \myloop \right)\\ \ +
\dots
\end{multline}
Each term, weighted by a factor $1/n!$, consists of all possible ways that the two external lines
with the full circles can connect each other through $n$ empty circles.
Note that the ``vacuum contribution'' factorizes out, leaving
\begin{equation} \label{twopoint}
\bra\p| \rho|\p'\ket\ = \  \Omega \left(\twopoint + \threepoint + \fourpoint +\dots\right)
\end{equation}
We call the two point function inside the parenthesis $K$:
\begin{equation}
\bra\p| \rho|\p'\ket\ \equiv \Omega \, K(\p, \p').
\end{equation}
When we consider the matrix elements in the $n$-particles sector, bubble diagrams again factorize out, leaving us with an expression of the form

\begin{equation}
    \bra\p_1 \p_2 \dots \p_n| \rho|\p_1' \p_2' \dots \p_n'\ket\ =\ \frac{\Omega}{n!}\sum{\left(2n-\rm points \; connected \; diagrams\right)} \ ,
\end{equation}
where now we are summing over all the possible diagrams that connect $n$ points on the left to $n$ points on the right. As we can have only two lines starting from each internal point, a diagram is composed by ``paths'', each of which connects one point on the left to one on the right (two points on the same side cannot be connected). This means that, if we select a pair of points $(\p_i,\, \p'_j)$, we can factorize an expression equal to the sum of connected diagrams in \eqref{twopoint}, which gives the two point function $K(\p_i, \p'_j)$. To obtain all the diagrams, we have to consider all the possible pairs; the result is that we can write all the matrix elements in terms of functions $K(\p_i, \p'_j)$:
\begin{equation}\label{term}
 \bra\p_1 \p_2 \dots \p_n| \rho|\p_1' \p_2' \dots \p_n'\ket\ =
 \frac{\Omega}{n!} \sum_{\sigma\in S_n} \prod_{j=1}^n K(\p_j, \p'_{\sigma(j)})\ .
\end{equation}
Note that this expression has the same structure as \eqref{element}, with $I_\beta(\x-\x')$ replaced by $K(\p, \p')$; the reason is that, using NW, local regions have the same Fock structure as the global space.



We define the powers of the two point function $K$ by multiplying $K$ as a one-particle operator acting
inside $P$. Accordingly, we define the trace of some power of $K$ as
\begin{equation}
{\rm Tr}\, K^m \ \equiv \ \int_{P^m} dp\, dp' \dots dp^{(m-1)}\, K(\p,\p')K(\p',\p'') \dots K(\p^{(m-1)},\p).
 \end{equation}
These traces are what we need  to calculate ${\rm Tr}_P \, \rho^n$ which, in turn, allows us to find Von Neumann entropy by means of eq. \ref{entropy} that can be applied to $\rho$ regardless of its normalization.

Consider first the case $n=2$; as $\rho$ is block diagonal in the fixed number of particles subspaces, so is $\rho^2$, and we can write, for the generic matrix element in the $m$-particles subspace
\begin{eqnarray}
    &&\bra\p_1 \p_2 \dots \p_m| \rho^2|\p_1' \p_2' \dots \p_m'\ket  \  \cr
    &&\qquad\qquad=\int_{P^m}d^d q_1  \dots d^d q_m \bra\p_1 \p_2 \dots \p_m| \rho| \q_1 \dots \q_m \ket
    \bra\q_1 \dots \q_m | \rho|\p_1' \p_2' \dots \p_m'\ket  \cr
    &&\qquad=
    \left(\frac{\Omega}{m!}\right)^2\int_{P^m}d^d q_1  \dots d^d q_m
    \sum_{\sigma\,,\sigma'\in S_m} \prod_{i\,j=1}^m K(\p_i, \q_{\sigma(i)})K(\q_j, \p'_{\sigma'(j)})   \cr
    &&\qquad=
    \left(\frac{\Omega}{m!}\right)^2\int_{P^m}d^d q_1  \dots d^d q_m
    \sum_{\sigma\,,\sigma'\in S_m} \prod_{j=1}^m K(\p_j, \q_{\sigma(j)})K(\q_{\sigma'(j)}, \p'_j)  \cr
    &&\qquad=
    \frac{\Omega^2}{m!}\sum_{\sigma\,\in S_m}
    \prod_{j=1}^m \int_{P}d^d qK(\p_j, \q)K(\q, \p'_{\sigma(j)}) \cr
    &&\qquad \equiv
    \frac{\Omega^2}{m!}\sum_{\sigma\,\in S_m}   \prod_{j=1}^m   K^2(\p_j, \p'_{\sigma(j)})\ .
\end{eqnarray}
Iterating this procedure we obtain the expression for $\rho^n$:
\begin{equation}
    \bra\p_1 \p_2 \dots \p_m| \rho^n|\p_1' \p_2' \dots \p_m'\ket \ =
    \frac{\Omega^n}{m!}\sum_{\sigma\,\in S_m}   \prod_{j=1}^m   K^n(\p_j, \p'_{\sigma(j)})\ .
\end{equation}
Finally, the trace is given by the sum of all the contributions of the $m$-particles matrix elements:
\begin{eqnarray} \label{rearrange}
   &&  {\rm Tr}_P \, \rho^n \ = \Omega^n\sum_{m=0}^{+\infty}\frac{1}{m!}\sum_{\sigma\,\in S_m}
    \int_{P^m}  d^d p_1  \dots d^d p_m \prod_{j=1}^m    K^n(\p_j, \p_{\sigma(j)})  \cr
   && \qquad\quad\ \ =\Omega^n \det (1-K^n)^{-1}
\end{eqnarray}
(see for example \cite{iz}, page 187, formula (4-86)), and then
\begin{equation} \label{exp}
{\rm Tr}_P \, \rho^n \ = \Omega^n \ \exp\left(\sum_{j=1}^{\infty}\frac{1}{j} {\rm Tr}\, K^{j n}\right)\, .
\end{equation}
Inserting this expression in \eqref{entropy}, we find
\begin{equation} \label{main}
S = \left. \left(-\frac{d}{d n} + 1\right)\sum_{j=1}^\infty \frac{1}{j}\, {\rm Tr}\,  K^{jn}\, \right|_{n=1}\ .
\end{equation}

The quantities we need to calculate are ${\rm Tr}\,  K^{n}$; we can give for these a diagrammatic expansion, like in \eqref{twopoint}. For $n=1$, we only have to ``close'' each diagram, so to match together the full circles at the ends of each factor, and we get

\begin{equation}
 {\rm Tr} K = \ \tadp \ +\ \looppr \ +\ \loopprr \ + \dots \;.
    \label{trk}
\end{equation}
Operator $K^2$ is obtained multiplying term by term two copies of the expression for $K$:

\begin{eqnarray}
   &&  K^2(\p,\p')= \left(\twopoint \ +\ \threepoint \ +\dots\right)\left(\twopoint \ + \ \threepoint \ +\dots\right)  \cr
   && = \lineppp \ \!\!\!\!\!\!\!\!\!\!\!\!\!\!\!\!\!\! + 2 \linepprp \ + \lineprprp \ \;\;\;\;\;\;\;\;\; +\dots \;;
\end{eqnarray}
when we take the trace, again we have to match the extremes of each diagram, so that we end up with closed loops with two full circles each and an arbitrary number of empty circles. The analogue expression for ${\rm Tr K^n}$ is a straightforward generalization: it contains loops with $n$ full circles and arbitrary empty circles. The explicit formula is
\begin{multline}
    {\rm Tr} K^n= \int d^dp_1\dots d^dp_n\sum^{+\infty}_{j_1\dots j_n=0}    \int d^dr'_1\dots d^dr'_{j_1} d^d r''_1 \dots \dots d^dr^{(n)}_1\dots d^dr^{(n)}_{j_n} \\[1.5mm]
    I_\beta(\p_1-\rr^1_1)
   I_\beta(\rr'_1-\rr^1_2)\dots I_\beta(\rr^1_{j_1}-\p_2) I_\beta(\p_2-\rr^2_1)\dots I_\beta(\rr^n_{j_n}-\p_1)
    \label{trkn} \; .
\end{multline}

\section{Explicit evaluations of entropy}
We consider separately the two situations $\beta^d \ll V$ (high temperature) and $\beta^d \gg V$ (low temperature), where $V$ is the volume of the region $P$ under consideration. We will mainly consider a massless field, but we will also consider a finite mass $m$ in the low temperature limit.

The high temperature limit, $V/\beta^d \rightarrow\infty$, is also the limit of very large volume, so we may expect to find, at leading order, the entropy that we would find by considering as our subsystem the whole space. Although this turns out to be the case, the actual proof is pretty
involved and is carried out in the subsection \ref{hightemp}.

\subsection{The Whole Space}

As a first check of our formalism we calculate the entropy of a system (not a subsystem) in a thermal state. This can be done by standard methods, {\it i.e.}, by calculating the partition function $Z={\rm Tr}\,e^{-\beta H}$ that, in our formalism, using eq. \eqref{element},  reads
\begin{equation} \label{xpartition}
Z=\sum^{+\infty}_{n=0}\int d^dx_1\dots d^dx_n\frac{1}{n!}\sum_{\sigma\in S_n}\prod_{j=1}^nI_\beta(\x_j-\x_{\sigma(j)})=\exp(\sum^{+\infty}_{j=1}\frac{1}{j}{\rm Tr}I_\beta^j)\; .
\end{equation}
We basically used the same derivation as for \eqref{exp}. In fact, in this case, $K = I_\beta$ since
there is nothing to integrate over outside the system.
The operator powers $I_\beta^j$ are obtained by integration over all space. For this purpose, we use eq. \eqref{iteration} and find
\begin{equation} \label{trace}
 {\rm Tr}I_{\beta}^n=\int d^dx I_{n\beta}(\x-\x)=VI_{n\beta}(0)\;.
\end{equation}
On the other hand, for $m=0$, we have
\begin{equation}
 I_{\beta}(0) = \int \frac{d^dk}{(2\pi)^d} e^{-\beta|\kk|}=
 \frac{\Omega_d}{(2\pi)^d} \int^{+\infty}_{0} dk k^{d-1}e^{-\beta k} = \frac{\Omega_d}{(2\pi)^d} \frac{(d-1)!}{\beta^d},
\end{equation}
where $ \Omega_d=2\pi^{d/2}/\Gamma(d/2)$ is the $d$-dimensional solid angle.
From \eqref{xpartition} we then find
\begin{equation}
    \ln Z = \frac{V}{\beta^d} \frac{(d-1)!\Omega_d}{(2\pi)^d}  \zeta(d+1),
\end{equation}
where $\zeta$ is the Riemann zeta function.
If we calculate the entropy by using
\begin{equation} \label{zentropy}
    S= \left(-\beta\frac{d}{d\beta}+1\right)\ln Z \;
\end{equation}
we find the leading order \eqref{hightentropy} of the more general result in the high temperature limit.

\subsection{High temperature limit} \label{hightemp}
We consider the limit $\beta^d/V \rightarrow 0$ in the massless case. Let us first consider the one dimensional case where $P$ is the the interval $(-L, L)$.
\subsubsection{One dimensional case}
We need to compute ${\rm Tr} K^n$. For $n=1$ it is given by (\ref{trk}). By eq. \eqref{1dim} the first term is simply
\begin{eqnarray}
 \ \tadp (\beta) =\int_{-L}^L \frac{\beta}{\pi (\beta^2 + (x-x)^2)}=\frac {2L}{\pi\beta}\ ,
\end{eqnarray}
where the term in parenthesis after the diagram, here and in what follows, specifies the suffix $\beta$ of the corresponding $I_\beta$ functions running in the loop. For the second term, we have
\begin{eqnarray}
&& \looppr (\beta,\beta)=\tadp (2\beta) -\looppp (\beta,\beta)=\nonumber \\[2mm]
&& \qquad\ \qquad\ \ \ =\frac {2L}{\pi 2\beta}\ -\int_{-L}^L dx \int_{-L}^L dy
\frac{\beta}{\pi (\beta^2 + (x-y)^2)}\frac{\beta}{\pi (\beta^2 + (y-x)^2)}\ .
\end{eqnarray}
In the first term of the latter equality the property \eqref{iteration} has been used. Now
\begin{eqnarray}
&& \int_{-L}^L dx \int_{-L}^L dy \frac{\beta}{\pi (\beta^2 + (x-y)^2)}\frac{\beta}{\pi (\beta^2 + (y-x)^2)}=\nonumber\\[1.5mm]
&& \qquad\qquad\ =-\frac {\beta}{2\pi^2} \frac d{d\beta} \int_{0}^{2L} dx \int_{0}^{2L} dy \frac{1}{\beta^2 + (x-y)^2}=\nonumber\\[1.5mm]
&& \qquad\qquad\ =-\frac {\beta}{\pi^2} \frac d{d\beta} \int_{0}^{2L} dx  \frac 1{\beta} \arctan \frac x\beta=
\frac 1{\beta\pi^2}\int_0^{2L} dx \frac d{dx} \left(x \arctan \frac x\beta \right)=\nonumber\\[1.5mm]
&& \qquad\qquad\ = \frac {2L}{\beta\pi^2} \arctan \frac {2L}\beta\ = \frac {L}{\pi\beta} +O(1)\ ,
\end{eqnarray}
thus
\begin{eqnarray}
\looppr=O(1)\ .
\end{eqnarray}
Thus it seems that terms containing external integrations give finite contributions. Before checking this to the next order, let us
note that
\begin{eqnarray}
&& \frac 1{\pi^{n+2}}\int_{-L}^L dx_0 \cdots \int_{-L}^L dx_{n+1} \frac {m_0 m_1 \cdots m_{n+1} \beta^{n+2}}{[\beta^2m_0^2+(x_0-x_1)^2]\cdots
[\beta^2 m_{n+1}^2+(x_{n+1}-x_0)^2]}=\cr
&& =\frac {2L}{\beta \pi^{n+2}}
\left. \int_{0}^\infty dx_1 \cdots \int_{0}^\infty dx_{n+1} \right[
\cr && \left. \frac {m_0 m_1 \cdots m_{n+1}}{[m_0^2+x_1^2][m_1^2+(x_1-x_2)^2]\cdots
[m_0^2+(x_n-x_{n+1})^2][m_{n+1}^2+x_{n+1}^2]} \right] \cr
&&\quad +\ {\rm cyclic\ terms}+\ldots \label{formula}
\end{eqnarray}
Here cyclic means with respect to the dependence on $m_i$ and the ellipses mean terms of the next order in $\beta/L$ (when $\beta/L$
goes to zero). This can be easily verified as follows:\\
Let us call it $U(m_0,\ldots,m_{n+1})$. One must show that the limit
\begin{eqnarray}
\lim_{\beta\rightarrow 0} \frac \beta{L} U(m_0,\ldots,m_{n+1})
\end{eqnarray}
exists and is given by the above expression. To this end, it is convenient first to rescale all coordinates by $\beta$,
so that
\begin{eqnarray}
&& \frac \beta{L} U(m_0,\ldots,m_{n+1})=\cr
&& \qquad\ =\frac 1{\pi^{n+2}} \frac \beta{L} \int_{-\frac L\beta}^{\frac {L}\beta} dx_0 \cdots \int_{-\frac {L}\beta}^{\frac {L}\beta}
dx_{n+1} \frac {m_0 m_1 \cdots m_{n+1}}{[m_0^2+(x_0-x_1)^2]\cdots [m_{n+1}^2+(x_{n+1}-x_0)^2]}.\label{quella}
\end{eqnarray}
Setting $z=L/\beta$, it suffices to compute the limit $z\rightarrow \infty$ by means of the de l'Hospital rule, to get the desired result.
Indeed the de l'Hospital rule says that we must look at the limit
\begin{eqnarray}
&& \lim_{z \rightarrow \infty}=\frac 1{\pi^{n+2}} \int_{-\frac L\beta}^{\frac {L}\beta} dx_1 \cdots \int_{-\frac {L}\beta}^{\frac {L}\beta}
dx_{n+1} \frac {m_0 m_1 \cdots m_{n+1}}{[m_1^2+(x_1-x_2)^2]\cdots [m_{n}^2+(x_{n}-x_{n+1})^2]}\cdot \cr
&& \qquad\ \!\!\! \left[ \frac 1{m_0^2+\left(\frac L\beta -x_1 \right)^2}\frac 1{m_{n+1}^2+\left(\frac L\beta -x_{n+1} \right)^2}
+\frac 1{m_0^2+\left(\frac L\beta +x_1 \right)^2}\frac 1{m_{n+1}^2+\left(\frac L\beta +x_{n+1} \right)^2} \right]\cr
&& \qquad\ +{\rm cyclic},
\end{eqnarray}
where we used the obvious relation
\begin{eqnarray}
\frac d{dz}\int_{-z}^z f(x)dx =f(z)+f(-z).
\end{eqnarray}
The two factors in the square brackets give the same contribution\footnote{the second one is obtained by the first one by changing sign to all
the integration variables}, so that after the shift $x_i\rightarrow x_i+L/\beta$, for all the $x_i$, our limit becomes
\begin{eqnarray}
&& \lim_{z \rightarrow \infty}=\frac 2{\pi^{n+2}} \int_{0}^{\frac
{2L}\beta} dx_1 \cdots \int_{0}^{\frac {2L}\beta} dx_{n+1} \frac
{m_0 m_1 \cdots m_{n+1}}{[m_1^2+(x_1-x_2)^2]\cdots
[m_{n}^2+(x_{n}-x_{n+1})^2]}\cdot \nonumber\\[1.5mm] && \qquad\ \cdot \frac
1{m_0^2+x_1^2}\frac 1{m_{n+1}^2+x_{n+1}^2}+{\rm cyclic}.
\label{questa}
\end{eqnarray}
It remains to show that indeed the integrals on the r.h.s. of (\ref{formula}) converge. This can be done by introducing the new variables
$t_i$ such that
\begin{eqnarray}
t_i=x_i -x_{i+1}\ , \quad i=1,\ldots, n \ , \quad t_{n+1}=x_{n+1}\ .
\end{eqnarray}
Then, after rewriting the integral, one sees that the integrand is dominated by $\prod_{i=0}^{n+1} \frac {m_i}{m_i^2+t_i^2}$.\\
Now, let us continue our analysis and consider the term
\begin{eqnarray}
\loopprr (\beta,\beta,\beta)=\tadp(3\beta)-\looppp(2\beta,\beta)-\looppp(\beta,2\beta)+\loopppp(\beta,\beta,\beta).
\end{eqnarray}
Using our general formula, find
\begin{eqnarray}
&& \tadp(3\beta)=\frac {2L}{3\pi\beta}+\ldots \nonumber\\[1.5mm]
&& \looppp(2\beta,\beta)=\looppp(\beta,2\beta)=\frac {8L}{\beta\pi^2}\int_{0}^{\infty} dx
\frac{1}{(1 + x^2)(4 + x^2)}+\ldots=\nonumber\\[1.5mm]
&&\qquad\qquad\ =\frac {8L}{3\pi^2\beta}\int_{0}^{\infty} dx\left[\frac 1{1+x^2}-\frac 1{4+x^2}\right]+\ldots
=\frac {2L}{3\pi\beta}+\ldots\cr
&& \loopppp(\beta,\beta,\beta)=\frac {6L}{\beta \pi^3}\int_0^\infty\!\! dx \int_0^\infty\!\! dy \frac 1{(1+x^2)(1+y^2)[1+(x-y)^2]}+\ldots
=\frac {2L}{3\pi\beta}\cr && \qquad\qquad\ +\ldots\ ,
\end{eqnarray}
so that
\begin{eqnarray}
\lim_{\beta\rightarrow 0}\ \left[ \beta \loopprr (\beta,\beta,\beta)\right]=0\ .
\end{eqnarray}
This is true for any power of $K$.
For example
\begin{eqnarray}
{\rm Tr} K^2 =\looppp +\loopppr +\loopprp +\ldots,
\end{eqnarray}
and using the above results we see that
\begin{eqnarray}
\loopprp(\beta,\beta,\beta)= \loopppr(\beta,\beta,\beta)=\looppp(\beta,2\beta)-\loopppp(\beta,\beta,\beta),
\end{eqnarray}
and then the $L/\beta$ terms drop out.\\
Indeed, one can prove that any loop integral containing at least a white ball insertion converges in the high energy limit. A complete
proof is very tedious, but tracks of a proof can be found in the Appendix.

Thus we get
\begin{eqnarray}
{\rm Tr} K^{n} =\looppn(n;\beta,\ldots,\beta)+\ldots,\label{dominant}
\end{eqnarray}
that is the leading contribution is given by the loop with exactly $n$ black ball insertions.
Now, being cyclic terms all coincident,
\begin{eqnarray}
&& \looppn(n;\beta,\ldots,\beta)=\nonumber\\[1.5mm]
&& \qquad =\frac {2nL}{\beta \pi^n}\int_{0}^\infty\!\! dx_1 \cdots \int_{0}^\infty\!\! dx_{n-1}
\frac 1{(1+x_1^2)(1+x_{n-1}^2)[1+(x_1-x_2)^2]\cdots[1+(x_{n-2}-x_{n-1})^2]}\nonumber\\[1.5mm]
&& \qquad +\ldots=\frac {2nL}{\beta \pi^n} U(n)+\ldots .
\end{eqnarray}
As we will see soon, the integral can be computed to give
\begin{eqnarray}
U(n)=\frac {\pi^{n-1}}{n^2}\ ,
\end{eqnarray}
so that
\begin{eqnarray}
&& \looppn(n;\beta,\ldots,\beta)=\frac {2L}{n\beta \pi}\ .\label{loop}
\end{eqnarray}
Indeed, we can show this as follows. Using the results shown in the Appendix, we can add a loop with a white insertion
and $n$ black insertions, without changing the divergent term, so that
\begin{eqnarray*}
&& \looppn(n;\beta,\ldots,\beta)\simeq\looppn(n;\beta,\ldots,\beta) +\looppnw(n;\beta,\ldots,\beta)
\\ && \qquad\qquad\ =\looppn(n-1;2\beta,\beta,\ldots,\beta) \ ,
\end{eqnarray*}
where $\simeq$ means equal up to convergent terms. Next we can add the term
\begin{eqnarray*}
\looppnw(n-1;2\beta,\beta,\ldots,\beta)
\end{eqnarray*}
to obtain
\begin{eqnarray*}
&& \looppn(n;\beta,\ldots,\beta)\simeq\looppn(n-2;3\beta,\beta,\ldots,\beta)\ .
\end{eqnarray*}
Proceeding in this way, we get
\begin{eqnarray}
&& \looppn(n;\beta,\ldots,\beta)\simeq\tadp(n\beta)\ , \label{integralone}
\end{eqnarray}
which gives (\ref{loop}).\\
Inserting (\ref{loop}) in (\ref{main}), we get for the entropy
\begin{eqnarray}
S=\frac {2\pi L}{3\beta}\ +\ldots\ .
\end{eqnarray}
Note that the above result is finite and extensive; no subtraction
needs to be made.

\subsubsection{The $d$ dimensional case}
The computations detailed out for the one dimensional case can be extended to any dimension $d\geq 1$.
Indeed, using (\ref{I}), one finds again that the loops containing only black insertions dominate so that
(\ref{dominant}) and (\ref{integralone}) continue to be true. On the other hand
\begin{eqnarray} \label{tracciakn}
\tadp(n\beta)=\frac {V}{(n\beta)^d} \frac {(d-1)!}{2^{d-1}\pi^{\frac d2} \Gamma(d/2)}+\ldots \ .
\end{eqnarray}
Thus we can easily compute the entropy, whose result is
\begin{eqnarray}\label{hightentropy}
S_d=\frac {V}{\beta^d} \frac {(d-1)!}{2^{d-1}\pi^{\frac d2} \Gamma(d/2)} (d+1) \zeta(d+1)+\ldots \ .
\end{eqnarray}
It can be useful to separate the odd dimensional cases from the even dimensional cases. Using the duplication formula for the Gamma function
one finds
\begin{eqnarray}
&& S_{2k} =\frac V{\beta^{2k}} \frac {(2k-1)!}{k! \pi^k 2^{2k+1}} (2k+1) \zeta(2k+1)+\ldots \ ,\nonumber \\[1.5mm]
&& S_{2k+1}=\frac V{\beta^{2k+1}} \frac {(k+1)!}{\pi^{k+1}} 2\zeta(2k+2) +\ldots \ .
\end{eqnarray}
For example, for $3$ spatial dimensions we find
\begin{eqnarray}
S_3=\frac V{\beta^3} \frac {2\pi^2}{45}+\ldots \ .
\end{eqnarray}


\subsection{Low temperature limits}

In this section we consider two different low temperature limits.
In the first case we take $\nu\rightarrow 0$, with
\begin{displaymath}
      \nu \equiv \frac{V}{(2 \pi \beta)^d} \; ,
\end{displaymath}
for fixed values of the product $\beta m$ between the inverse temperature and the mass of the field. This can be thought as
the small volume limit at fixed values of mass and temperature. However, we can also interpret it as the low temperature limit
of a massless field; thus the mass plays the role of a IR regulator so that successively the quantity $\beta m$ must be set to zero.

In the second case, we consider the limit $\beta \rightarrow \infty$ for fixed values of the volume $V$ and the mass $m$, that is the low temperature
limit of a massive field.

\subsubsection{Small volumes and massless limit}

Looking at $\nu\rightarrow 0$ as a small volume limit, the result may depend on the shape of $P$ and the way it ``shrinks''.
To avoid this problem, we will suppose for $P$ to shrink down isotropically. This means that for any $m$ dimensional
section of $P$, having volume $V_m$, the quantity $V_m/\beta^m$ must tend to zero when $\nu\rightarrow 0$.
Note that such condition is automatically satisfied in the low temperature limit interpretation, if $P$ is
contained in  a compact region.
For integrals over regions shrinking down isotropically, we can then use the approximation\footnote{this is the form of the spatial integrals
after a substitution $\p\rightarrow \p/\beta$; the $\beta^d$ extra-factors are reabsorbed by rescaling the momentum variables as
$\kk\rightarrow \beta\kk$.}
\begin{equation} \label{approx}
\int_{\frac{P}{\beta^d}}d^dp f(\p) = (2 \pi)^d \nu f(0) +{\cal O}(\nu^2).
\end{equation}
To estimate ${\rm Tr}K^n$ we must approximate all loop integrals. The simplest one, is the tadpole integral:
\begin{equation}
    \tadp= \int_{P}d^dpI_{\beta}(\p -\p)\, = \, VI_{\beta}(0)  \,  =\,
    \nu\int d^dk e^{-\sqrt{\kk^2 +(\beta m)^2}}\;.
\end{equation}
 Using \eqref{approx}, we see that each integration over $P$ can be simplified giving a factor $\nu$. For example, for the two-points loop we get
\begin{eqnarray}
    && \looppp = \int_{P\times P}d^dpd^dq I_{\beta}(\p -\q)I_{\beta}(\q-\p)   \  \cr
    && = \frac{1}{(2 \pi)^{2d}}\int d^dkd^dk' \int_{\left(\frac{P}{\beta^d}\right)^2}d^dpd^dq
    e^{-i(\kk-\kp)(\p-\q)-\left(\sqrt{\kk^2 +(\beta m)^2}+\sqrt{\kp^2 +(\beta m)^2}\right)} \ \cr
    && = \nu^2 \left(\int d^dk e^{-\sqrt{\kk^2 +(\beta m)^2}} \right)^2 + {\cal O}(\nu^3)\; .
\end{eqnarray}
Note that the last integral depends on the temperature only via the fixed product $\beta m$, so that the resulting expression
is of order $\nu^2$. In the same way we conclude that the loop with $n$ full circles is of order $\nu^n$.
Diagrams with empty circles can be computed using the identity
\begin{equation}
    \int_Rd^dr=\int d^dx - \int_{P}d^dp \; .
\end{equation}
The integrals over the whole space can be reduced using \eqref{crucial}-\eqref{iteration}, whereas each integral over $P$ gives
a contribution proportional to $\nu$, which is therefore a subleading term with respect to the first one.
For example, for the ``mixed'' two-points loop we get
\begin{multline}
    \looppr \, =\,  \int d^dx\int_{P}d^dpI_{\beta}(\p-\x)I_{\beta}(\x-\p)\, -\,  \looppp =\nonumber\\[1.5mm]
     \int_{P}d^dpI_{2\beta}(0) + {\cal O}(\nu^2) \,  =\,
   \nu \frac{\Omega_d}{2^d}\int^{+\infty}_{0} dk k^{d-1}e^{-\sqrt{k^2+(2\beta m)^2}} + {\cal O}(\nu^2)\; .
\end{multline}
In conclusion, to estimate the generic loop integral can simply remove each empty circle by means of the substitution
$\int_Rd^dr I_{\beta}(\p-\rr)I_{\tilde \beta}(\rr-\q)\rightarrow I_{\beta+\tilde \beta}(\p-\q)$,
so that the leading order of each diagram is determined by the number of its full balls.\\
By applying this procedure to all terms in \eqref{trk}, we find
\begin{equation}
    {\rm Tr}K = V \sum^{+\infty}_{j=1} I_{j\beta}(0)+ {\cal O}(\nu^2)=
     \nu\Omega_d\sum^{+\infty}_{j=1} \frac{1}{j^d}\int^{+\infty}_{0} dk k^{d-1}e^{-\sqrt{k^2+(j\beta m)^2}} + {\cal O}(\nu^2) =
    \nu \Omega_d C_d + {\cal O}(\nu^2)\; ,
\end{equation}
where $C_d$ is a multiplicative factor that doesn't depend on $\nu$.

The diagrammatic expansion of ${\rm Tr} K^n$ contains $n$ full circles at any loop, so that for the leading term we get
\begin{eqnarray} \label{tracciaklow}
    && {\rm Tr} K^n=\sum^{+\infty}_{j_1\dots j_n=0}\int d^dp_1\dots d^dp_n
    I_{j_1\beta}(\p_1-\p_2)\dots I_{j_n\beta}(\p_n-\p_1)    + {\cal O}(\nu^{n+1})= \cr
    && \prod^{n}_{l=1}\left(\sum^{+\infty}_{j_l=0}\nu I_{j_l\beta}(0)\right)+{\cal O}(\nu^{n+1})=
    \left(\nu\Omega_d C_d\right)^n+{\cal O}(\nu^{n+1}) \; .
\end{eqnarray}
Plugging this result into \eqref{drops} and then into \eqref{entropy}, we find for the entropy
\begin{equation}
    S = - \nu\, \Omega_d C_d \ln  \nu \, +\,  {\cal O}(\nu) \; .
    \label{lowent}
\end{equation}
Note that, for $d>1$, $C_d$ is a finite number. Indeed, it is a function of $\beta m$ defined by the series
\begin{equation}
    C_d = \sum^{+\infty}_{j=1} \frac{1}{j^d}\int^{+\infty}_{0} dk k^{d-1}e^{-\sqrt{k^2+(j\beta m)^2}} \; ,
    \label{serie}
\end{equation}
whose terms are positive and bounded by the terms of the converging series $C_d(0)$. Thus
\begin{displaymath}
    C_d(\beta m)  \,  \leq \, C_d( 0) \, =\, \sum^{+\infty}_{j=1} \frac{1}{j^d}\int^{+\infty}_{0} dk k^{d-1}e^{-k}\, = \, (d-1)!\zeta(d)\; .
\end{displaymath}
An IR divergence appears in the onedimensional massless case. Obviously it could be cured
by an IR cutoff that limits the integrations to the external region $R$.
However, a natural regularization is provided by the mass term. Indeed, we can fix $\beta m$ at arbitrarily small but positive values, so that
the general term in the series \eqref{serie} satisfies
\begin{eqnarray}
  && \int^{+\infty}_{0} dk e^{-\sqrt{k^2+(j\beta m)^2}} =
 \int^{j\beta m}_{0} dk e^{-\sqrt{k^2+(j\beta m)^2}} + \int^{+\infty}_{j\beta m} dk e^{-\sqrt{k^2+(j\beta m)^2}} \ \cr
    && \leq e^{-j\beta m}j\beta m + \int^{+\infty}_{j\beta m} dk e^{-k} = e^{-j\beta m}(j\beta m + 1)\; ,
\end{eqnarray}
ensuring convergence.
For $d>1$, no divergences occur and in the massless case the entropy \eqref{lowent} reads
\begin{eqnarray}
    && S \sim -X\ln X \;, \cr
    && X = \frac{V}{\beta^d}2^d\pi^{\frac{d-1}{2}}\Gamma(\frac{d+1}{2})\zeta(d)\,  \qquad (d>1\, ,\; \beta m =0) \; .
    \label{lowmassless}
\end{eqnarray}
Again, the above result is finite and no subtractions need to be made.

\subsubsection{Low temperatures for a massive field} \label{lowmassive}
Let us first consider the one dimensional case. For $\beta\rightarrow\infty$, the two point function $I_{\beta}(x)$ behave as
\begin{equation}
    I_{\beta}(x) \sim \sqrt{\frac{m}{2\pi\beta}}e^{-m\sqrt{\beta^2+x^2}}\; .
    \label{asintotico}
\end{equation}
When $x$ falls inside $P$, the whole expression can be approximated by a constant:
\begin{displaymath}
    I_{\beta}(p) \sim I_{\beta}(0) \sim \sqrt{\frac{m}{2\pi\beta}}e^{-m\beta}\; \;,
\end{displaymath}
and each integration over $P$ contributes with a term proportional to the volume; thus we get
\begin{eqnarray}
 && \tadp = \int^{L}_{-L} dp I_{\beta}(p-p)=2L\,I_{\beta}(0)\sim 2L\sqrt{\frac{m}{2\pi\beta}}e^{-m\beta} \;,\cr
 && \looppn(n\, {\rm points}) \sim \int^{L}_{-L} dp_1\dots\int^{L}_{-L} dp_n\left(\,I_{\beta}(0)\right)^n
 \sim \left(2L\sqrt{\frac{m}{2\pi\beta}}e^{-m\beta}\right)^n    \; .
\end{eqnarray}
It follows that a loop with $n+1$ integrations over $P$ is subleading with respect to one with $n$ integrations and we can
use again the usual trick to substitute integrations over $R$ with integrations over the whole space, finding
\begin{displaymath}
    {\rm Tr} K \sim L\sum^{+\infty}_{n=1}I_{n\beta}(0)\;.
\end{displaymath}
Now, each addend $I_{n\beta}(0)$ is suppressed by a factor $\left(e^{-m\beta}\right)^n$, so that only the first term in the sum is relevant
when $\beta\rightarrow\infty$. The same argument can be applied to ${\rm Tr} K^n$, giving
\begin{eqnarray}
 && {\rm Tr} K^n \sim \left(2L\sqrt{\frac{m}{2\pi\beta}}e^{-m\beta}\right)^n    \; ,
\end{eqnarray}
from which we get the following expression for the entropy
\begin{eqnarray}
    && S \sim -X\ln X \;, \cr
    && X = 2L\sqrt{\frac{m}{2\pi\beta}}\,e^{-m\beta}\; .
    \label{lowmassived1}
\end{eqnarray}
These analysis can be easily extended to any dimension. Indeed, the two point function $I_{\beta}(p)$ is still suppressed by a factor $e^{-m\beta}$
for $p\in P$, so that all computation work essentially in the same way, providing for the entropy an expression of the form \eqref{lowmassived1},
where the exponential multiplies a function of $\beta$ with an at most polynomial growth. We will not enter into more details.

\section{Conclusions}
In this paper we have considered the problem of
computing the entropy of a subsystem confined in a finite volume region
of a quantum field theory system. As hinted in the introduction, the subsystem under consideration is taken to be deep inside a much larger system which has reached thermal equilibrium and is therefore described by a thermal Gibbs state.
The entropy of the subsystem results to be divergent in a non standard but curable way. The divergent part is the entanglement entropy of the vacuum and is a function of a cut-off and of the area of the surface bounding the region. Here we have highlighted the general issue of assigning appropriate quantum degrees of freedom to the considered region. Indeed, the inside/outside separation is traditionally realized through the usual
localization prescription, which attributes to that region the local relativistic fields therein defined, together with their conjugated momenta. This in fact leads to a tensor product separation (TPS) of the whole Hilbert space, but it must face the above described problem of the infinities in the calculation of the entropy.

In this paper we have adopted the Newton-Wigner (NW) localization scheme to define the TPS of the quantum fields associated to classical spatial separation. We made use of the creation and annihilation operators which act on the vacuum generating the NW
position eigenvectors. At first, one could feel disturbed by the fact that these
are non relativistically invariant operators. However, this does not affect the
relativistic characterization of the dynamics. It simply means that localization
is not observer independent. After all, a place/region of space is always to be defined on some chosen $t = constant$ hypersurface. Moreover, Lorentz invariance is naturally broken
by thermalization. In the NW prescription the vacuum is a product state, so that the
corresponding divergent, area-dependent contribution to the entropy never
appears in the calculations.
We argue that the NW prescription is the
right one in order to treat the problem of coarse-graining a thermalized
microscopic system. The first point we bring to support our thesis is the
finiteness of the results we obtained using such a prescription. We worked with
a Klein-Gordon scalar field on a flat Minkowski spacetime in arbitrary dimensions,
and computed the Von Neumann entropy for a subsystem confined in a region of volume $V$ at a finite
temperature $T$. The NW prescription automatically regularizes the ultraviolet divergences and directly gives finite results. This permitted us to analyze various
peculiar situations.

In the high temperature/large volume limit we
immediately obtained the expected thermodynamic result:
entropy is extensive and, for a massless field, goes as $VT^d$, where $V$ is the volume of the region and $d$ the number of spatial dimensions. At leading order our results match the standard calculation of the entropy of a field system with appropriate conditions at its boundary \cite{kapusta}.
In other words, in the NW approach, a generic subsystem of a larger thermal system is also approximately thermal and has the same temperature.
Again, this seems to suggest that NW localization is more appropriate for the spatial coarse-graining of microscopic quantities.

At low temperature/small volume, thermal
entropy in our calculation goes to zero, but, differently from the standard
approach, no unusual subtraction is needed.
In the canonical approach
\cite{cala,ads} thermal entropy is sub-extensive, $S_{\rm therm}\simeq (VT^d)^{(d+1)/d}$ at low temperature, whereas our regularized entropy
approaches extensivity from above ($S\simeq -VT^d \ln VT^d$ for small $VT^d$). Since the calculations in the two approaches follow two completely
different routes, it is difficult to recognize the technical reason for this discrepancy. Note, however, that at very low temperatures the modes
that are typically excited have wavelengths much larger than the size of the subsystem itself. Such modes are not
contemplated in the spectrum of the locally defined Hamiltonian, whose lowest non-zero eigenvalues are of order $\sim1/V^{1/d}$. Those long-wavelength
correlations between the internal and the external region dominate in this limit and appear to be the cause of our super-extensive entropy. Note also
that $-\epsilon \ln \epsilon$ is the generic asymptotic behavior of the Von Neumann entropy of a density matrix with a parameter $\epsilon$ which
becomes a pure state in the limit $\epsilon \rightarrow 0$.

Our renormalization procedure, although unconventional, looks encouraging.
The thermodynamical description of quantum field systems is a lively subject, which finds
its application beyond the physics of complex systems, going through a better understanding
of interconnections between gravitational and quantum effects, as for example black hole
thermodynamics or AdS/CFT correspondence. However, many of such interesting applications
are affected by the problem of the ultraviolet divergences so that, apart from some exceptional cases,
many results remain at a qualitative or conjectural level. If holographic entropy bounds (e.g. \cite{vene,bousso}) and the area-dependent black hole entropy (see \cite{damour}, and also \cite{solo,myers} for entropy renormalization in that context) are to be taken as meaningful signals of quantum gravity effects,
one may want to consistently get rid of the comparable area dependent contribution that appears already at low energies in flat space.
Because of the automatic
regularization, the finiteness of our approach seems to provide a powerful and concrete computational method to overcome such technical empasses.

\section*{Acknowledgments}
We enjoyed useful conversations and correspondence with Francesco Belgiorno,
Horacio Casini, Xiao Liu, Lee Smolin and Andrew Tolley. Research at Perimeter
Institute is supported by the Government of Canada through Industry Canada and by the Province of
Ontario through the Ministry of Research and Innovation.

\appendix

\section{Tracks of a proof}
Let us consider the integral
\begin{eqnarray}
&& J= \int_{R_0} dx_0 \cdots \int_{R_{n+1}} dx_{n+1} \frac {m_0 m_1 \cdots m_{n+1} }{[m_0^2+(x_0-x_1)^2]\cdots
[ m_{n+1}^2+(x_{n+1}-x_0)^2]}, \label{sperem}
\end{eqnarray}
where the ranges $R_i$ can be the interval $I_L=[-L/\beta,L/\beta]$ or the set $E_L=(-\infty,-L/\beta]\cup [L/\beta,\infty)$, and $n\geq 1$ (the case $n=0$
can be verified by a direct computation). Without loss of generality we can assume $R_{n+1}=E_L$. We can perform the integration in $dx_{n+1}$ by
means of the formula
\begin{eqnarray}
&& \int dx \frac {ab}{[a^2+(x-y)^2][b^2+(z-x)^2]} =\cr
&& \qquad\ =\frac {b[b^2-a^2+(z-y)^2]\arctan \frac {x-y}a+a[a^2-b^2+(z-y)^2]\arctan \frac {x-z}b}{[(y-z)^2+(a+b)^2][(y-z)^2+(a-b)^2]}\cr
&& \qquad\ -\frac {ab(y-z)\log\frac {a^2+(x-y)^2}{b^2+(x-z)^2}}{[(y-z)^2+(a+b)^2][(y-z)^2+(a-b)^2]},
\end{eqnarray}
so that
\begin{eqnarray}
&& W(m_n,m_{n+1};x_n,x_{0};\beta)=\int_{E_L} dx_{n+1} \frac {m_n m_{n+1}}{[m_n^2+(x_n-x_{n+1})^2][m_{n+1}^2+(x_{n+1}-x_0)^2]}=\cr
&& \qquad\ =  -\frac {m_{n+1}[m_{n+1}^2-m_n^2+(x_{0}-x_n)^2]\arctan \frac {\frac L\beta-x_n}{m_n}}{[(x_n-x_{0})^2+(m_n+m_{n+1})^2][(x_n-x_{0})^2+(m_n-m_{n+1})^2]}\cr
&& \qquad\ -\frac {m_n[m_n^2-m_{n+1}^2+(x_{0}-x_n)^2]\arctan \frac {\frac L\beta-x_{0}}{m_{n+1}}}{[(x_n-x_{0})^2+(m_n+m_{n+1})^2][(x_n-x_{0})^2+(m_n-m_{n+1})^2]}\cr
&& \qquad\ +\frac {m_n m_{n+1}(x_n-x_{0})\log\frac {m_n^2+(\frac L\beta-x_n)^2}{m_{n+1}^2+(\frac L\beta-x_{0})^2}}{[(x_n-x_{0})^2+(m_n+m_{n+1})^2][(x_n-x_{0})^2+(m_n-m_{n+1})^2]}\cr
&& \qquad\ -(\beta\rightarrow -\beta) +
\frac{\pi (m_{n+1}+m_n)[(m_{n+1}-m_n)^2 +(x_0-x_{n})^2]}{[(x_n-x_{0})^2+(m_n+m_{n+1})^2][(x_n-x_{0})^2+(m_n-m_{n+1})^2]} .
\end{eqnarray}
Next, a very careful analysis is needed, distinguishing the cases if $R_0$, $R_n$ are of type $I_L$ and/or $E_L$. Note that it is an odd function of
$\beta$, apart from a term which do not contain $\beta$. Indeed such term imply $W(m_n,m_{n+1};x_n,x_{0};\beta)\rightarrow 0$ if
$\beta \rightarrow 0^+$.
A very lengthy computation shows that one can find a set of positive constants $K_{ab}$ such that
\begin{eqnarray}
W(a,b;x_n,x_0;\beta)    \leq \frac {K_{ab}}{(x_n-x_{0})^2+(a+b)^2} \frac \beta{L}.
\end{eqnarray}
After substitution in (\ref{sperem}), we see that in the worst case\footnote{that is when all the remaining insertions are the black ones}
$J$ takes the form (\ref{quella}) so that the limit $\beta\rightarrow 0^+$ exists.

\end{document}